\documentclass{article}
\usepackage{fullpage}
\usepackage{spconf,amsmath,graphicx}
\usepackage{caption}
\usepackage{subcaption}
\usepackage{amsfonts}
\usepackage{booktabs}
\usepackage{adjustbox}
\usepackage{amsfonts}

\usepackage{multirow,multicol, makecell, booktabs}
\usepackage{amssymb}% http://ctan.org/pkg/amssymb
\usepackage{pifont}% http://ctan.org/pkg/pifont
\newcommand{\cmark}{\ding{51}}%
\newcommand{\xmark}{\ding{55}}%
\usepackage{longtable}
\newlength\mylength
\newcolumntype{C}[1]{>{\centering\arraybackslash}p{#1}}

\makeatletter
\newcommand*\titleheader[1]{\gdef\@titleheader{#1}}
\AtBeginDocument{%
  \let\st@red@title\@title
  \def\@title{%
    \bgroup\normalfont\large\centering\@titleheader\par\egroup
    \vskip1.5em\st@red@title}
}
\makeatother
\titleheader{© 2021 IEEE.  Personal use of this material is permitted.  Permission from IEEE must be obtained for all other uses, in any current or future media, including reprinting/republishing this material for advertising or promotional purposes, creating new collective works, for resale or redistribution to servers or lists, or reuse of any copyrighted component of this work in other works.}

\title{A Lightweight ReLU-Based Feature Fusion for Aerial Scene Classification}
\name{Md Adnan Arefeen$^{\star}$ \qquad Sumaiya Tabassum Nimi$^{\star}$ \qquad Md Yusuf Sarwar Uddin$^{\star}$ \qquad Zhu Li$^{\star}$}
  
  \address{$^{\star}$ Department of EECS, University of Missouri-Kansas City, MO, USA}
\begin{document}
\topmargin=0mm 
%\ninept
%
\maketitle
\begin{abstract}
In this paper, we propose a transfer-learning based model construction technique for the aerial scene classification problem. The core of our technique is a layer selection strategy, 
named ReLU-Based Feature Fusion (RBFF), 
that extracts feature maps from a pretrained CNN-based single-object image classification model, namely MobileNetV2, and constructs a model for the aerial scene classification task. RBFF stacks features extracted from the batch normalization layer of a few selected blocks of MobileNetV2, where the candidate blocks are selected based on the
characteristics of the ReLU activation layers present in those blocks. The feature vector is then compressed into a low-dimensional feature space using dimension reduction algorithms on which we
train a low-cost SVM classifier for the classification of the aerial images. We validate our choice of selected features based on the significance of the extracted features with respect to our classification pipeline. RBFF remarkably does not involve any training of the base CNN model except for a few parameters for the classifier, which makes the technique very cost-effective for practical deployments. The constructed model despite being lightweight outperforms
several recently proposed models in terms of accuracy
for a number of aerial scene datasets.

\end{abstract}
\section{Introduction}
\vspace{-3mm}
Aerial image classification involves categorizing aerial images into an exhaustive set of known classes based on the contents of the images. 
%With the abundance of high-resolution images captured by different apparatus for earth observation, the demand for automatic identification and labeling of these images is growing. 
In recent years, several research works have been conducted to classify remote sensing images because of its application in the military and civil tasks such as disaster management \cite{kyrkou2019deep}, traffic supervision \cite{koga2018cnn}, %urban planning \cite{cao2018urban}, 
environmental monitoring \cite{wang2019accurate} and urban mapping~\cite{wu2020deepdualmapper}. The earliest efforts towards aerial image classification employed pixel-level and object-level analysis approaches~\cite{ cheng2015effective}. 
In the recent years, deep learning models such as Convolutional Neural Networks (CNN) have emerged as one of the best tools for computer vision tasks like image classification and face recognition. Consequently, CNN models have been applied for aerial image classification also, leading to better classification performance compared to the prior approaches. 
\begin{figure}[!htpb]
    \centering
    \includegraphics[width=0.48\textwidth]{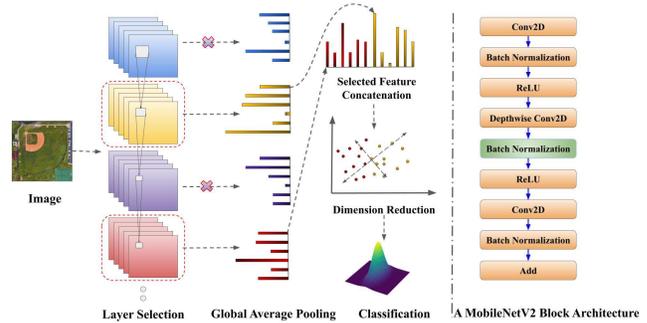}
    \caption{ReLU-Based Feature Fusion in MobilenetV2 Block Architecture}
    \label{fig:model}
\end{figure}
Most of the research works that employ CNN based approaches for the task of remote scene classification train the CNN model from scratch on the pertinent dataset~\cite{zhang2020multi, bi2020multiple, bi2020radc}. The training phase involved in these approaches is computationally very expensive.\\ 
In the current work, we propose a transfer learning-based methodology for the same task (i.e., remote scene classification) that \emph{does not} require training of any CNN model. 
%We use the strength of transfer learning with a notion of using the same model for two different tasks i.e. to classify the ImageNet dataset objects and aerial images simultaneously. 
Our proposed approach, ReLU-Based Feature Fusion (RBFF), involves stacking features extracted from batch normalization layers of a few \emph{selected} blocks based on a layer significance value from the lightweight MobileNetV2~\cite{sandler2018mobilenetv2} model pretrained on ImageNet~\cite{imagenet_cvpr09} dataset. A block of MobileNetV2 with the proposed pipeline is depicted in Figure~\ref{fig:model}.
%We use MobileNetV2 as our base model because it is one of the most lightweight deep learning models designed till date. 
%The selection of blocks from the MobileNetV2 model is done based on a novel strategy called ReLU-Based Feature Fusion (RBFF) that . 
Before concatenating the extracted features, we use the global context of the features which helps reducing the dimension of the feature maps without eliminating the global context of the features. We further compress the features using dimension reduction techniques such as Principal Component Analysis (PCA)~\cite{wold1987principal} and Linear Discriminant Analysis (LDA)~\cite{lachenbruch1979discriminant} so that we get a more compact low-dimensional feature space without major loss of information. Finally, we use a one-vs-rest support vector machine (SVM) model on the compressed feature space for classification of the aerial images. The cost-effectiveness of our proposed model is two-fold. The model is space-efficient as its size is very small compared to the deep learning models that have been proposed so far for the task of remote scene classification ($3.28\sim13.28$ MB). %Also, the associated training of the model is very minimal. We mostly transfer deep features from a pretrained model instead of training the entire model from scratch. Hence, our approach does not incur the associated training overhead.
Furthermore, we investigate the accuracy vs model size trade-off issue as our technique employs two dimension reduction techniques in the pipeline, which comes with additional overhead in terms of the overall model size. 

\begin{figure}
    \centering
    \includegraphics[width=0.9\linewidth]{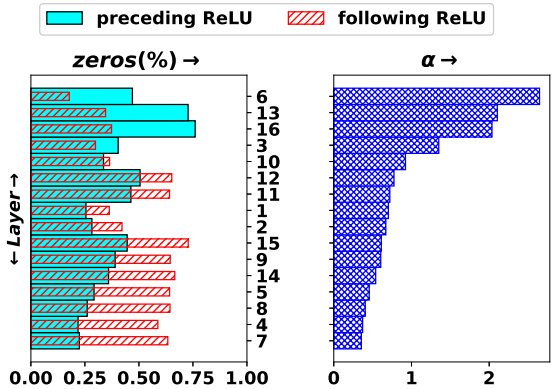}
    \caption{Zero percentage for the preceding ReLU layer and the following ReLU layer for each layer of AID dataset (left). The layers are sorted according to highest value of $\alpha$ (right).}
    \label{fig:zero_percentage VS alpha}
\end{figure}
\vspace{-2mm}
\section{Background and Motivation}
\vspace{-3mm}
We have used pretrained lightweight MobilenetV2 model to extract the deep features through a layer selection strategy (RBFF), Whereas in \cite{cao2020self}, a nonparametric self-attention layer is proposed to get the more discriminative features from the pretrained CNN model (VGG-VD16 \cite{krizhevsky2012imagenet} and AlexNet \cite{simonyan2014very}). In~\cite{cheng2018deep}, novel loss function was proposed so that the feature space of the deep learning model learned on the objective would discern between images representing separate classes, so that features of the images belonging to the same class would be mapped closer in the learned feature space and vice-versa for images belonging to different classes. %For this approach to work, the model would have to be trained from scratch using proposed objective function. 
Meanwhile, in~\cite{anwer2018binary}, texture information of images obtained using Local Binary Pattern (LBP) was used in addition to traditional RGB images for training deep learning models. %It was shown shown that fusing this additional information, both early and late into the network pipeline resulted in better performance on both texture recognition and aerial mage classification. 
%We note again that, for this approach, the network needed to to be trained from scratch with the additional texture information. 
On the contrary, in~\cite{bi2020radc}, a novel network architecture was proposed for aerial image classification, that consisted of dense blocks with addition and modification of residual attention layers and a classification layer. The network was designed to contain fewer parameters than the traditional models. %But this approach also incorporated overhead of training deep learning models from scratch. 
Traditional VGGNet architecture was modified in~\cite{guo2019global} to incorporate local and global attention layers in order to capture the corresponding information from aerial images, along with subsidiary objectives added to the loss function for additional supervision.
%This approach was also not free from the overhead of from the scratch model training. We note that, 
\par Contrary to all these approaches, our proposed method is free from the huge overhead of deep learning model training. To classify aerial images, we follow a layer selection based transfer learning based approach where we apply a novel technique to select the most informative layers from a pretrained model. 
\section{Methods}
\vspace{-3mm}
%\subsection{Motivation}
%\vspace{-3mm}
\subsection{ReLU-Based Feature Fusion}
\begin{figure}

  \begin{subfigure}[b]{0.48\columnwidth}
    \includegraphics[width=\linewidth]{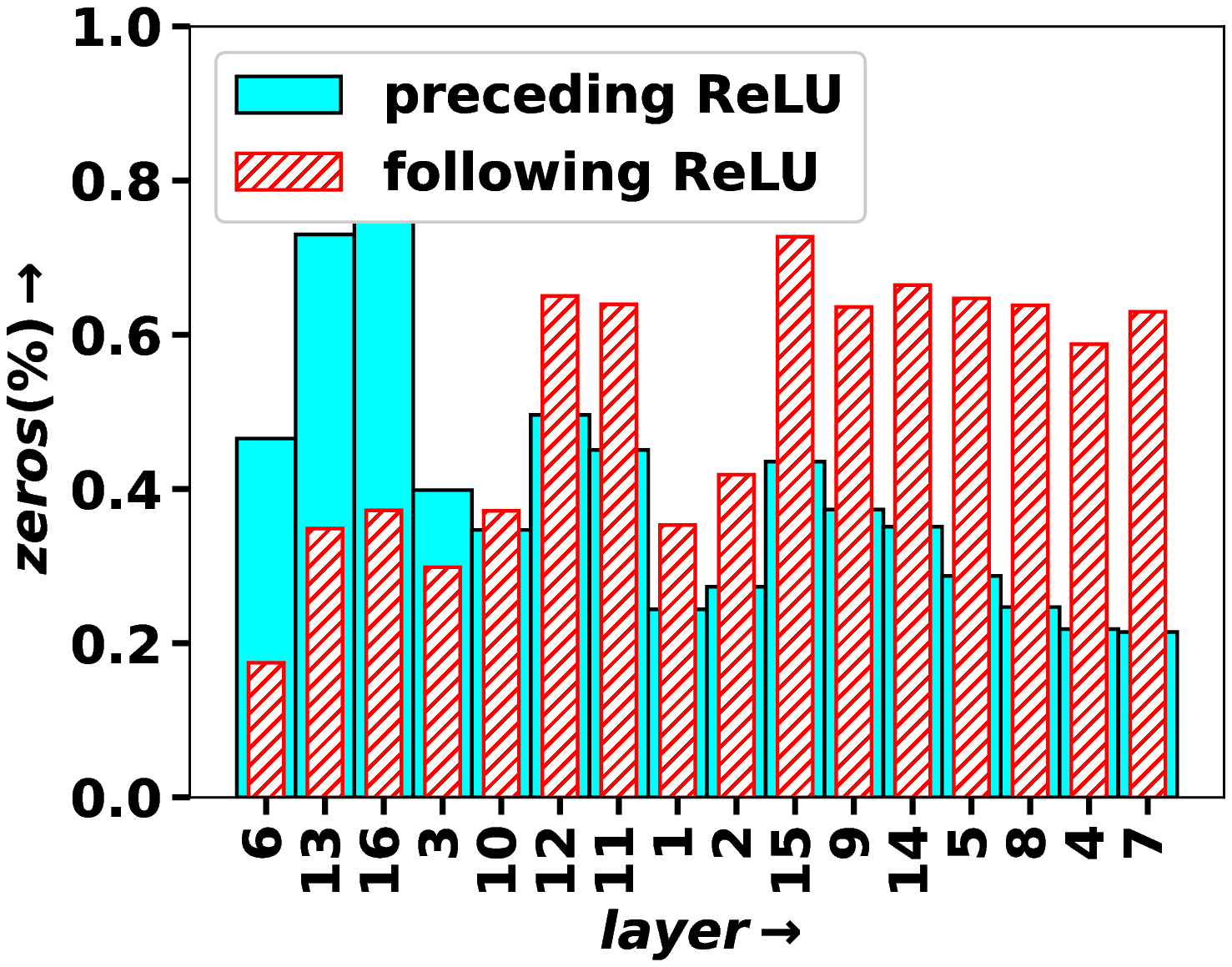}
    \caption{NWPU}
    \label{fig:zero-nwpu}
  \end{subfigure}
  \begin{subfigure}[b]{0.48\columnwidth}
    \includegraphics[width=\linewidth]{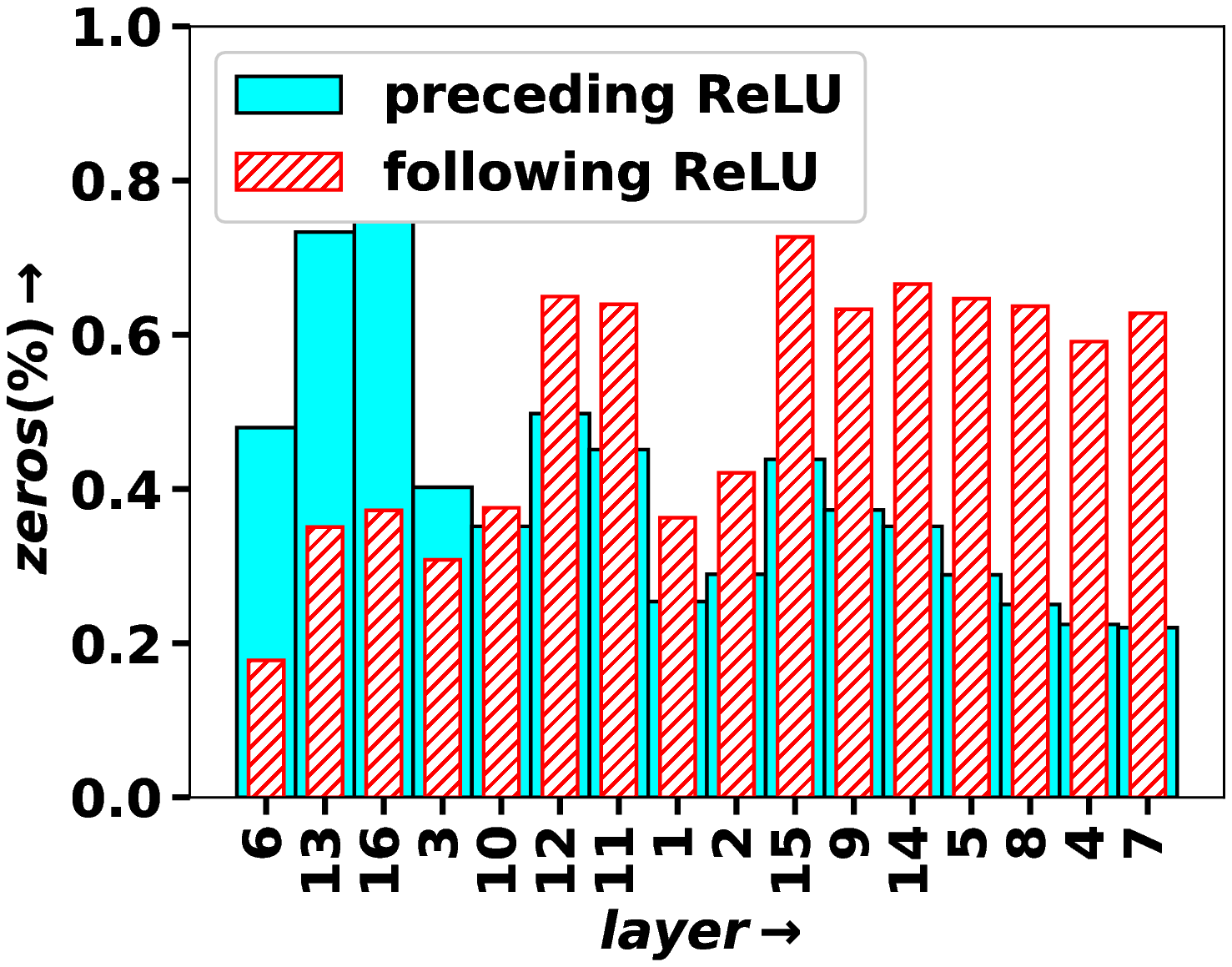}
    \caption{UCM}
    \label{fig:zero-ucm}
  \end{subfigure}
  \caption{Zero percentage for the preceding ReLU layer and the following ReLU layer for each block.%For each dataset except the white image, 5 samples from each class are chosen to calculate the zero percentage after these two layers. Then the average zero percentage is computed from the available zero percentage value for each sample of a dataset.
  }
  \label{fig:zero-percentage}
\end{figure}

\vspace{-3mm}
%As described, we extract features from the pretrained MobileNetV2 model for the aerial scene classification task. One trivial approach can be 
%Extracting features from \emph{all} layers will obviously result in accumulating a large number of features for a single image. Instead, we choose a subset of layers from the base model.
The most trivial approach can be the extraction of features from each layer which will result in accumulating a large number of features for a single image. Hence, rather than extracting features from each layer, we intuitively choose a subset of layers from the MobileNetV2 architecture for feature extraction. 
%First, we select four blocks for feature extraction in an iterative manner and observe the accuracy. For MobileNetV2, the more we go further, the larger the number of features as the number of channels increases in the later layers than the layers at the beginning. We observe that, 
In MobileNetV2~\cite{sandler2018mobilenetv2}, each block architecture is of ReLU $\circ$ Conv $\circ$ BN $\circ$ ReLU like fashion. There are
%We apply this characteristic of the block to estimate the significance of a layer. 
 $16$ of such blocks in MobileNetV2. One such block is depicted in Figure~\ref{fig:model}. 
%To accomplish the feature fusion task, we observe the characteristic of the middle BN (Batch Normalization) layer (marked with a different color in Figure~\ref{fig:model}). A more intuitive explanation behind choosing this layer is discussed in the next section. 
The BN layer within a block is surrounded by two ReLU layers in the same block, which are referred to as \emph{preceding} and \emph{following} ReLU layers respectively with respect to that BN layer. The ReLU layers are activation layers that replace its input $x$ to $\max(0,x)$.
%If the preceding ReLU layer of a given BN layer collapses more zeros than the following ReLU layer, we infer that the BN layer is an important one for feature extraction because less information is lost through the forward propagation after that BN layer. Consequently, 
The ratio of the fraction of zeros in the \emph{preceding} and \emph{following} ReLU layers around a BN layer designates the degree of significance of the BN layer in a given block. We denote this significance by $\alpha$ for each block. More formally, the layer significance, $\alpha$, per block is defined as: 

%In this paper, we have used the term ($\alpha$) to indicate the significance of a BN layer. We consider the preceding and following ReLU layer to measure the ($\alpha$). The layer significance $\alpha$ is defined as,
\iffalse
\begin{equation}
\begin{split}
    \alpha &= \frac{Z_{prev}}{Z_{next}}\\
    Z &= \frac{1}{N}\sum_{i=1}^{N} (1 - v_i) \\
    v_{i} &= \frac{\sum_{k=1}^{n} \mathbb{I} [P>0]}{\sum_{k=1}^{n} P}
   % \frac{1}{\alpha} &= \frac{V_{prior}}{V_{posterior}}\\
%     &= \frac{1 - Z_{prior}}{1 - Z_{posterior}}\\
\end{split}
\label{eqn:prepost}
\end{equation}
\fi
\begin{align*}
%\begin{split}
    \alpha &= \frac{Z_{prev}}{Z_{next}}; \nonumber&&
    Z_{(.)}&= \frac{1}{N}\sum_{i=1}^{N} (1 - v_i); \nonumber &&
    %v_{i} &= \frac{\sum_{k=1}^{n} \mathbb{I} [\boldsymbol{U}_{jk}>0]}{\sum_{k=1}^{n} |\boldsymbol{U}_{jk}|}
    v_{i} &= \frac{ \sum_{x\in \boldsymbol{U}^i}x > 0}{ |\boldsymbol{U}^i|}
    %v_{i} &= \frac{ \sum \boldsymbol{U}^i > 0}{ |\boldsymbol{U}^i|}
   % \frac{1}{\alpha} &= \frac{V_{prior}}{V_{posterior}}\\
%     &= \frac{1 - Z_{prior}}{1 - Z_{posterior}}\\
%\end{split}
\label{eqn:prepost}&&
\end{align*}
\noindent
where $Z_{prev}$ and $Z_{next}$ denote the average zero volume percentage of the preceding and the following ReLU layers of a BN layer, respectively. By zero volume, we mean how many zero elements exist in the output after a ReLU layer. If $Z_{prev} > Z_{next}$ i.e. $\alpha > 1$, the corresponding BN layer is more informative and vice versa. For an output tensor of $i^{th}$ ReLU layer, $\boldsymbol{U}^i \in {\rm I\!R}^{\mathcal{H} \times \mathcal{W} \times \mathcal{ C}}$, where $\mathcal{H}$, $\mathcal{W}$, and $\mathcal{C}$ refers to height, width, and channels of the feature map, we compute positive volume percentage of $i^{th}$ layer $v_i$ for each input image being fed to the network. So, $(1-v_i)$ is the zero volume percentage of a layer. To ensure minimal training, we randomly take $5$ images from each class, calculate $v_i$ of the layer for each image, and average them to get the general view of the layer. In this way, we compute the $Z(.)$ for a given layer. The average positive volume of a layer is reciprocal to the average percent of zeros in a layer in terms of $\alpha$. Conceptually, $\alpha$ measures the importance of a batch normalization layer of each block. The higher the value of $\alpha$ is, the more important the batch normalization layer is in terms of its suitability to be included in the feature extraction. Consequently, we stack features from top BN layers that have higher $\alpha$ values. Figure \ref{fig:zero_percentage VS alpha} shows the significance of BN layers for all $16$ blocks of MobilenetV2 for AID dataset. The layers are sorted with respect to higher value of $\alpha$. We observe from the figure that, the top discriminative blocks are $3$, $6$, $13$, and $16$ (the blocks with higher $\alpha$ values for their respective BN layers).
Similar characteristic is also identified for NWPU and UCM dataset in Figure~\ref{fig:zero-percentage}.
%Figure \ref{fig:zero-nwpu} shows similar picture for NWPU dataset. We note that the impact of the percentage of zeros is almost the same across all datasets, which we recognize as the characteristic of the base network. \\To verify this even further, we feed a \emph{white} image, that is, an image having all one values in it, to the network and observed the same characteristic (Figure~\ref{fig:zero-ucm}).
\\Based on the $\alpha$ values, we stack features from the BN layers of blocks 3, 6, 13, and 16 of MobileNetV2 (these are the blocks that have the largest $\alpha$ values). To extract the global context information of the features from the selected layers, we apply the global attention based feature extraction. 
The global average pooling (GAP)~\cite{lin2013network} is used to achieve this. %GAP is used extensively in the literature to reduce the spatial 3-D dimensional tensor. 
For a channel $c$ of $i^{th}$ layer feature map, $\boldsymbol{U}^i_c$
%If an output tensor of a ReLU layer, i.e. a feature map, is denoted by $\textbf{U} \in {\rm I\!R}^{\mathcal{H} \times \mathcal{W} \times \mathcal{ C}}$, where $\mathcal{H}$, $\mathcal{W}$, and $\mathcal{C}$ refers to height, width, and channels of the feature map, then for a channel $c$, 
the global average pooling will be,

\begin{equation}
    GAP(\boldsymbol{U}^i_c) = \frac{1}{H \times W} \sum_{j=1}^{H} \sum_{k=1}^{W} u^i_{c} [j,k]
\end{equation}
For each channel the 2-D $H \times W$ feature map is converted to one value. Thus, global average pooling captures the global information of each channel of a layer output by converting 3-D tensor to a series of vector. By this way, a global attention based feature map is achieved.

\subsection{Batch Normalization Layer in the Context of LDA}
\vspace{-3mm}
%Once the features are extracted from the set of selected layers of the base model and the global average pooling is applied on the stacked features, we apply dimension reduction techniques on the flattened feature vector. We use the well-known principal component analysis (PCA) and linear discriminant analysis (LDA) for this. In an $n$-class classification problem, LDA reduces the dimension of feature space to $k$ ($k \leq n-1$), so that the separation between the features belonging to different classes is maximized. Furthermore, LDA works with the assumption that the data is Gaussian. And this is why we take out the feature maps obtained from the Batch Normalization (BN) layers as that makes this assumption to hold. As we see in~\cite{bjorck2018understanding}, batch normalization enforces the activations to be normally distributed by a shift factor $\beta$ and a scale factor $\gamma$. 
\iffalse
\begin{equation}
    BN(x) = \frac{x_{in} - \mu_B}{\sqrt{\sigma^2_B + \epsilon}};
    x_{out} = \gamma BN(x) + \beta
\end{equation}
\fi
%Here, $x_{in}$ and $x_{out}$ are input and output of BN layer. The mean and standard deviation of input activation over a mini-batch is denoted by $\mu$ and $\sigma$ respectively. It is also discussed the paper that without batch normalization, activations tend to be heavy-tailed. However, 
Batch normalization repeatedly enforces the activations to have zero mean and unit standard deviation~\cite{bjorck2018understanding}, leading to faster and optimal convergence of the neural network training. In our approach, we use this fact to our advantage. Prior to the SVM classification, we use LDA for dimension reduction. Since LDA works best for normally distributed data, our choice of using the feature maps from the BN layers leads to the optimal performance of LDA and hence the optimal classification of the aerial images. 
%Before performing LDA, we have also used principal component analysis (PCA) to reduce the feature dimension is a two-stage fashion so that the redundancy is eliminated from the output feature map. This two-stage feature reduction strategy sometimes is costly compared to the actual base model. We discuss this issue in Section~\ref{sec:model_sizevsacc}.

%In the subsection ``Model size vs Accuracy'' of Section Results, we have discussed this issue.

\section{Results}
%\vspace{-4mm}
\subsection{Datasets}
\vspace{-3mm}
\begin{table}[]
\caption{Description of Datasets}
\label{tab:dataset}
\centering
    \adjustbox{width=0.8\linewidth}{
\begin{tabular}{ccc}
\toprule[1.5pt]
Dataset    & Number of Classes & Number of Samples per Class \\
\midrule[1.2pt]
UCM        & 21                & 100                         \\
AID        & 30                & 220-400                     \\
NWPU       & 45                & 700                         \\
%PatternNet & 38                & 800                        \\
\bottomrule[1.5pt]
\end{tabular}
}
\end{table}
In this paper, we have used three widely used aerial scene datasets, namely UCM~\cite{yang2012geographic}, AID~\cite{xia2017aid}, NWPU~\cite{cheng2017remote} %and PatternNet~\cite{zhou2018patternnet}.
The description of these datasets in terms of the number of classes and the number of samples per class is listed in Table \ref{tab:dataset}.
 \begin{table}[]
%\large
\caption{Model Parameters and Feature Dimension}
\label{tab:model-params-feature-dims}
\centering

\adjustbox{width=0.95\linewidth}{
\begin{tabular}{cccc}
\toprule[1.5pt]
Model     & \multicolumn{1}{p{2cm}}{\centering Parameters \\(in Millions)} & \multicolumn{1}{p{2cm}}{\centering Model Size \\(in MB)} & Feature Dimension  \\
\midrule[1.5pt]

AlexNet~\cite{xia2017aid}   & 60  & 434  & 4096          \\ 
VGG-VD-16~\cite{xia2017aid} & 138 & 1024 & 4096             \\ 
GoogLeNet~\cite{xia2017aid} & 6.8 & 91.1 & 1024           \\ 
RADC-Net~\cite{bi2020radc}  & 0.5 & 9.0 & 352            \\
MIDC-Net\_CS~\cite{bi2020multiple} & 0.5 & 10.0 & --\\ 
RBFF (3\_6\_13\_16) &  1.46 & \textbf{6.28} & 1872 \\
RBFF (3\_6\_13) &  0.59 & \textbf{2.73} & 912 \\
RBFF + PCA + LDA + SVM & - & 3.28$\sim$13.28 & $k$ - 1\\
\bottomrule[1.5pt]
\end{tabular}
}
\end{table}
\subsection{Experimental Settings}
\vspace{-3mm}
We consider the BN layers of the top 3 and 4 blocks among 16 blocks in the descending order of their $\alpha$ values as evident from Figure~\ref{fig:zero_percentage VS alpha}. After the feature extraction phase, we successively perform GAP, PCA, LDA, and SVM on the features and report the classification results using cross-validation technique. 
\begin{table*}[]

\caption{Accuracy comparison of RBFF with Different Methods for Different Datasets}
    \centering
    \label{tab:new_all}
    \adjustbox{width=\linewidth}{
    \begin{tabular}{c|cc|cc|cc}
    \toprule[1.5pt]
    Method & AID (20\%) & AID (50\%) & NWPU (10\%) & NWPU (20\%) & UCM (50\%) & UCM (80\%)\\
    \midrule[1.2pt]
    PLSA (SIFT) \cite{xia2017aid} & 56.24 $\pm$ 0.58 & 63.07 $\pm$ 1.77 & -- & -- & 67.55 $\pm$ 1.11 & 71.38 $\pm$ 1.77\\
    BoVW (SIFT) \cite{xia2017aid} & 62.49 $\pm$ 0.53 & 68.37 $\pm$ 0.40 & 41.72 $\pm$ 0.21 & 44.97 $\pm$ 0.28 & 73.48 $\pm$ 1.39 & 75.52 $\pm$ 2.13\\
    LDA (SIFT) \cite{xia2017aid} & 51.73 $\pm$ 0.73 & 68.96 $\pm$ 0.58 & -- & -- & 59.24 $\pm$ 1.66 & 75.98 $\pm$ 1.60\\
    AlexNet \cite{xia2017aid} &  86.86 $\pm$ 0.47 & 89.53 $\pm$ 0.31 &76.69 $\pm$ 0.21 & 79.85 $\pm$ 0.13 & 93.98 $\pm$ 0.67 & 95.02 $\pm$ 0.81\\
    VGGNet-16 \cite{xia2017aid} & 86.59 $\pm$ 0.29 & 89.64 $\pm$ 0.36 & 76.47 $\pm$ 0.18 & 79.79 $\pm$ 0.15 & 94.14 $\pm$ 0.69 & 95.21 $\pm$ 1.20\\
    GoogLenet \cite{xia2017aid}  & 83.44 $\pm$ 0.40 & 86.39 $\pm$ 0.55 & 76.19 $\pm$ 0.38 & 78.48 $\pm$ 0.26 & 92.70 $\pm$ 0.60 & 94.31 $\pm$ 0.89\\
    SPP with Alexnet \cite{han2017pre} & 87.44 $\pm$ 0.45 & 91.45 $\pm$ 0.38 & 82.13 $\pm$ 0.30 & 84.64 $\pm$ 0.23 & 94.77 $\pm$ 0.46 & 96.67 $\pm$ 0.94\\
    D-CNN with AlexNet \cite{cheng2018deep} & 85.62 $\pm$ 0.10 & \textbf{94.47} $\pm$ 0.12 & 85.56 $\pm$ 0.20 & 87.24 $\pm$ 0.12 & -- & 96.67 $\pm$ 0.10\\
    TEX-Net with VGG \cite{anwer2018binary}  & 87.32 $\pm$ 0.37 & 90.00 $\pm$ 0.33 & -- & -- & 94.22 $\pm$ 0.50 & 95.31 $\pm$ 0.69\\
    Gated Attention \cite{ilse2018attention} & 87.63 $\pm$ 0.44 & 92.01 $\pm$ 0.21 & 84.94 $\pm$ 0.22 & 86.62 $\pm$ 0.22 & 94.64 $\pm$ 0.43 & 96.12 $\pm$ 0.42\\
    MIDC-Net\_CS \cite{bi2020multiple}   & 88.51 $\pm$ 0.41    & 92.95 $\pm$ 0.17 & \textbf{86.12} $\pm$ 0.29    & 87.99 $\pm$ 0.18 & 95.41 $\pm$ 0.40    & \textbf{97.40 $\pm$ 0.48}\\
    RADC-Net \cite{bi2020radc}   & 88.12 $\pm$ 0.43 & 92.35 $\pm$ 0.19 & 85.72 $\pm$ 0.25 & 87.63 $\pm$ 0.28 & 94.79 $\pm$ 0.42 & 97.05 $\pm$ 0.48\\
    VGG\_VD16 + SAFF \cite{cao2020self}   & 90.25 $\pm$ 0.29 &  93.83 $\pm$ 0.28 & 84.38 $\pm$ 0.19 &  87.86 $\pm$ 0.14 & -- &  97.02 $\pm$ 0.78\\

    \midrule[1.2pt]
    %RBFF(3\_6\_13\_16) + PCA (600) + LDA + SVM (Ours)       & \textbf{91.01} $\pm$ 0.20 & 93.07 $\pm$ 0.18 \\
    %RBFF(3\_6\_13) + PCA + LDA + SVM (Ours)       & \textbf{90.95} $\pm$ 0.21 & 93.46 $\pm$ 0.39 \\
    %RBFF + PCA + LDA + SVM (Ours) ($600\times600\times3$)      & \textbf{92.01} $\pm$ 0.26 & 94.07 $\pm$ 0.33 \\
    %RBFF(3\_6\_13\_16) + LDA + SVM (Ours)  & 26.30 $\pm$ 2.01 & 93.64 $\pm$ 0.48\\
    %RBFF(3\_6\_13) + LDA + SVM (Ours) & 88.67 $\pm$ 0.21 & 93.64 $\pm$ 0.42\\

    RBFF (3\_6\_13) + PCA (600) + LDA + SVM  & \textbf{90.94 $\pm$ 0.26} & 93.73 $\pm$ 0.23 & 84.59 $\pm$ 0.24 & 87.53 $\pm$ 0.19 & 94.69 $\pm$ 0.34 & 96.40 $\pm$ 0.99 \\
    %RBFF (3\_6\_13) + PCA (600) + LDA + SVM (600 $\times$ 600 $\times$ 3)  & \textbf{91.21 $\pm$ 0.26} & 93.81 $\pm$ 0.31 & -- & -- & -- & -- \\
    %RBFF (3\_6\_13)  + PCA (900) + LDA + SVM  & 88.78 $\pm$ 0.3 & 93.66 $\pm$ 0.41 & 83.46 $\pm$ 0.25 & 87.55 $\pm$ 0.25 & 80.1 $\pm$ 0.3 & 94.88 $\pm$ 1.05 \\
    RBFF (3\_6\_13\_16) + PCA (600) + LDA + SVM  & \textbf{91.02 $\pm$ 0.22} & 92.90 $\pm$ 0.31 & 84.24 $\pm$ 0.29 & 86.53 $\pm$ 0.20 & \textbf{95.83 $\pm$ 0.54} & 96.90 $\pm$ 1.14 \\
    %RBFF (3\_6\_13\_16) + PCA (600) + LDA + SVM (600 $\times$ 600 $\times$ 3)  & \textbf{91.72 $\pm$ 0.27} & 94.02 $\pm$ 0.39  & -- & -- & -- & -- \\
    %RBFF (3\_6\_13\_16) + PCA (1000) + LDA + SVM  & \textbf{90.44 $\pm$ 0.15} & 93.6 $\pm$ 0.12 & 84.2 $\pm$ 0.27 & 87.26 $\pm$ 0.22 & 92.55 $\pm$ 0.76 & 96.93 $\pm$ 0.76 \\
    
    RBFF (3\_6\_13) + LDA + SVM   & -- & 93.64  $\pm$ 0.42 & -- & 87.5 $\pm$ 0.24 & -- & 94.93 $\pm$ 0.82\\
    RBFF (3\_6\_13\_16) + LDA + SVM  & -- & 93.64 $\pm$ 0.48 & -- & \textbf{88.05} $\pm$ \textbf{0.18} & -- & 82.38 $\pm$ 1.92\\
    %RBFF (3\_6\_13\_16) + LDA + SVM  & 27.59 $\pm$ 1.8 & 93.91 $\pm$ 0.15 & --  & \textbf{88.05} $\pm$ \textbf{0.18} & 92.88 $\pm$ 0.98 & 82.38 $\pm$ 1.92\\

    \bottomrule[1.5pt]
    \end{tabular}
    }

\end{table*}
%\noindent
The size of the actual MobileNetV2 model is approximately $14$ MB. After transfer learning using our proposed RBFF technique, the model size is reduced to $6.28$ MB as shown in Table~\ref{tab:model-params-feature-dims}. Excluding the last layer (layer 16) from RBFF, the size is only $2.73$ MB. In Table~\ref{tab:model-params-feature-dims}, $k$ means the number of classes as after LDA, the features are reduced to only number of classes minus one if explicit reduction is not provided to the LDA model.  We choose 600 as the number of components for PCA. The reason is discussed in section~\ref{sec:model_sizevsacc}. We also show results without PCA to perform the ablation study. The detail is discussed in section~\ref{sec:ablation}. These experiments are done under a fixed random state for PCA, RepeatedStratified K-Fold, and SVM. The random states are 3, 33, and 333 respectively for PCA, RepeatedStratified K-Fold, and SVM to reproduce the results. We have used the {\tt scikit-learn} package~\cite{scikit-learn} to implement these algorithms. Our experiments have been done on NVIDIA Quadro RTX 4000 GPU and i7-9750 CPU@2.60GHz.
\begin{figure}
\centering
    \centering

%\begin{subfigure}[b]{0.33\textwidth}
%\centering
%    \includegraphics[width=0.9\textwidth,height=4cm]{Images/model_size_v1.png}
%    \caption{}
%    \label{fig:model-size}
%\end{subfigure}
%
\begin{subfigure}[b]{0.48\linewidth}
\centering
    \includegraphics[width=\textwidth]{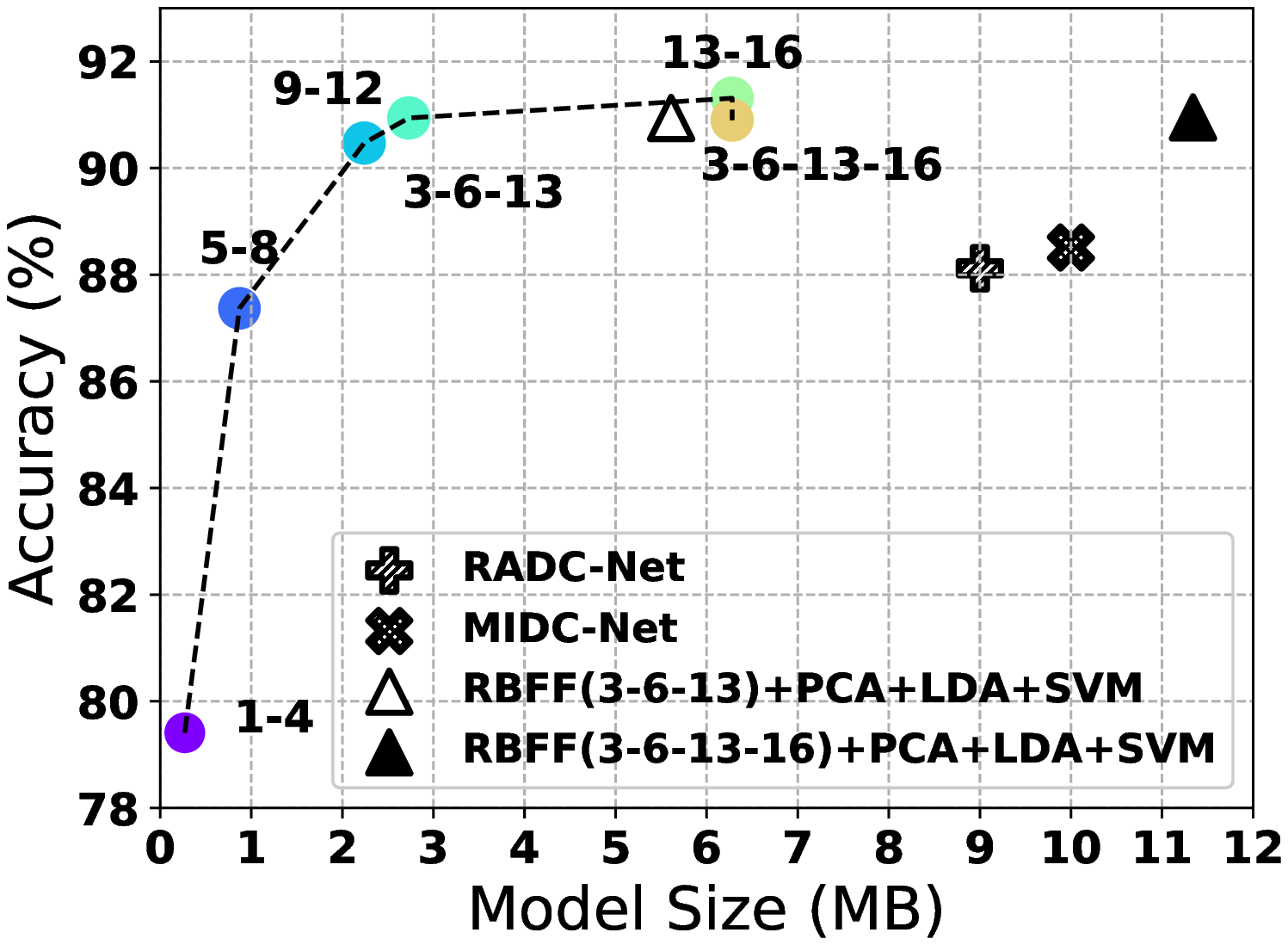}
    \caption{}
  \label{fig:model-error-rate}
\end{subfigure}
\begin{subfigure}[b]{0.50\linewidth}
    \centering
    \includegraphics[width=\textwidth]{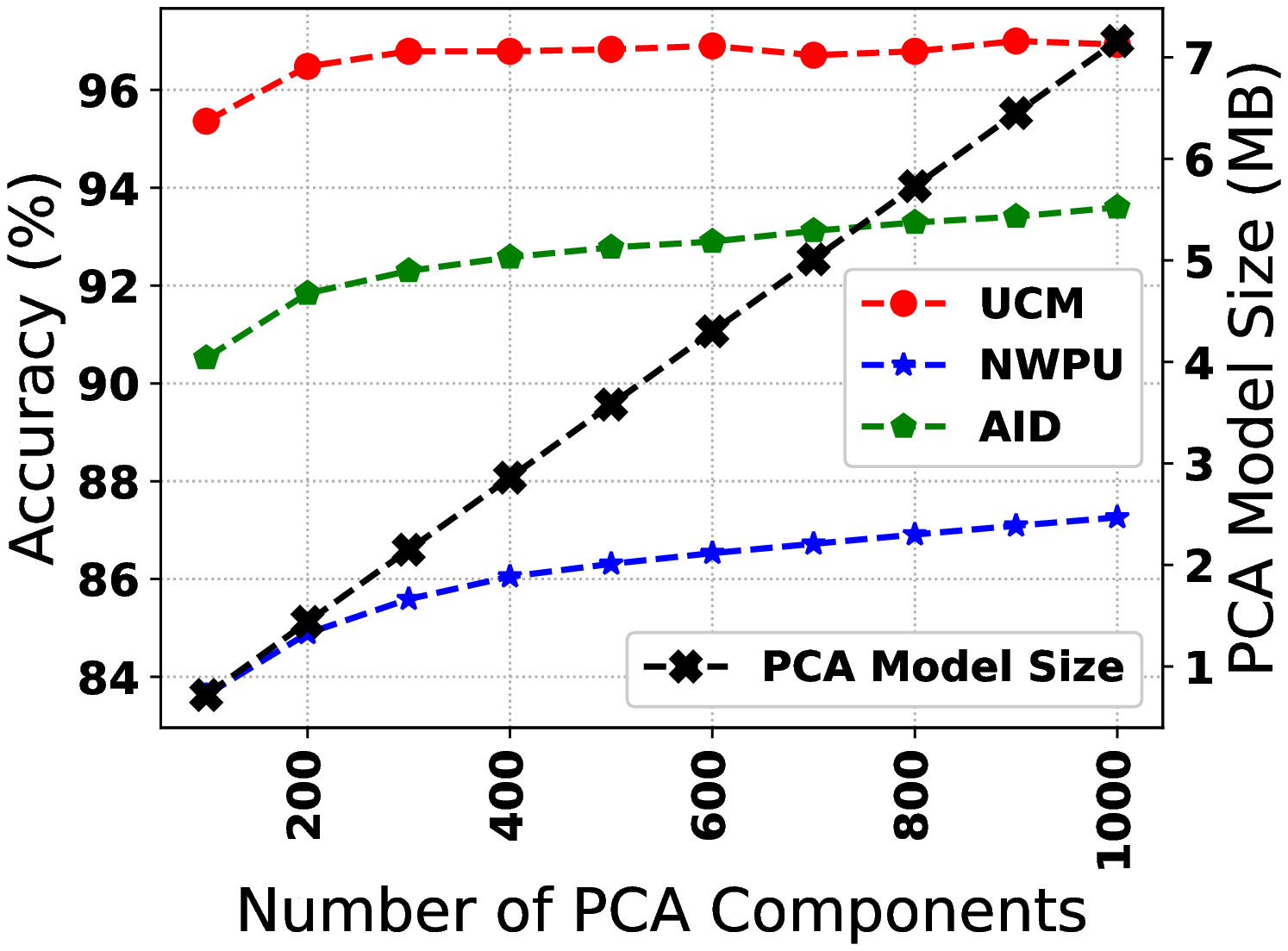}
    \caption{}
    \label{fig:pca_model_size_vs_accuracy}
\end{subfigure}
    \caption{%(a) Comparison of model size with various chunks of blocks on AID dataset. $20\%$ of the data is used as training set for each k-fold cross validation. 
    (a) Comparison of sorted mean accuracy with total model size for the same dataset (AID ($20\%$)) and same configuration. (b) Accuracy, model size and the number of PCA components for all three datasets used in this paper.
    }
    \label{fig:my_label}

\end{figure}
\subsection{Experimental Analysis}
\vspace{-3mm}
We compare our approach with some recently proposed network architectures in terms of classification accuracy. Table~\ref{tab:new_all} represents the overall accuracy results for three datasets. The AID and NWPU datasets are comparatively harder to classify than the UCM dataset. During experimentation, the images are resized to $224 \times 224$ and preprocessed using the predefined {\tt preprocess\_input} function of MobileNetV2 application of Keras for ImageNet dataset. %We also experimented with the image dimension $600 \times 600$ for the AID dataset and reported the results. 
For AID with $20\%$ training dataset, our proposed approach outperforms all the state-of-the-art results. With regard to the $50\%$  training set, D-CNN with Alexnet outperforms all the results. RBFF holds the second best result. %We observe that the inclusion of PCA in the pipeline contributes towards better classification performance except for the NWPU dataset for $20\%$ training case of NWPU dataset. 
For $20\%$ training set of the NWPU dataset and $50\%$ for UCM, RBFF outperforms all recent works. For other cases, RBFF is only less than $0.5-1.53\%$ than the best results. 
%and For $50\%$ for UCM, it outperforms all recent works. 

%For the PatternNet dataset, a little variation in accuracy is observed only after the decimal point. Table~\ref{tab:experiment} shows the results for all datasets in different combinations in terms of training size and layer selection strategy.
\begin{table}[]
\caption{Comparison of Accuracy with Layer selection} %for AID (50\% training)}
\label{tab:all3layers}
    \centering
    \adjustbox{width=0.70\linewidth}{
    \begin{tabular}{ccc}
    \toprule[2pt]
    Layer & Accuracy & Feature Dimension \\
    \midrule[1.5pt]
    9\_10\_11 & 92.18 $\pm$ 0.27 & 1344\\
    9\_10\_12 & 92.81 $\pm$ 0.39  & 1344\\
    10\_11\_12 & 92.48 $\pm$ 0.13 & 1536\\
    9\_11\_12 & 92.37 $\pm$ 0.33 & 1536 \\
    3\_6\_13 (Ours)& \textbf{93.64 $\pm$ 0.42} & \textbf{912}\\
    
    \bottomrule[2 pt]
    \end{tabular}
}
\end{table}
\vspace{-3mm}
\subsection{Model size vs Accuracy}\label{sec:model_sizevsacc}
\vspace{-3mm}
In Figure~\ref{fig:model-error-rate}, we show the superioirity of RBFF to other lightweight models~\cite{bi2020multiple,bi2020radc}. To examine how the model size changes if we choose different depthwise BN layers, we divide the $16$ selected BN layers into four chunks, namely 1--4, 5--8, 9--12, and 13--16 and observe accuracy vs model size (Figure~\ref{fig:model-error-rate}). %Figure~\ref{fig:model-error-rate} shows how the model size increases with the increase of layer chunks, along with the associated increase in accuracy. 
As layers 9--12 exhibit  similar performance compared our proposed RBFF, we compare the 3--6--13 with each three combinations from 9--12 for AID ($50\%$ training) in Table~\ref{tab:all3layers} where RBFF has a lower cost with higher accuracy.
The increase of the PCA model size with the increase of PCA components is illustrated in Figure~\ref{fig:pca_model_size_vs_accuracy}. Based on Figure~\ref{fig:pca_model_size_vs_accuracy}, we have chosen 600 components for PCA for having higher accuracy and comparatively lower PCA model size. One can use this hyperparameter as a knob to fit when necessary.
%As per MobileNetV2 architecture, layers 1--4 have fewer features than layers 13--16.

%In Figure~\ref{fig:model-error-rate}, we observe how the accuracy increases if we take features from deeper layers. Notably, the model size increases if we consider the deeper layers for transfer learning as the deep layers are larger in dimension than their shallow counterparts. The increase in feature dimension plays a crucial role in the enhancement of the PCA model size as illustrated in Figure~\ref{fig:model-size}. For layers 13--16, although accuracy is higher compared to other choices of layers, the feature dimension and total model size are also higher.

%\section{Case Study}
%\subsection{PCA components vs Accuracy vs PCA model size}
%Now we analyze how the number of PCA components affects the classification accuracy. We use all 4 datasets for the analysis. From the reported percentage of training samples, the highest ones are used for each dataset. For each dataset, we extract features from layers 3, 6, 13, and 16 using the RBFF technique, increase PCA components by a number of 100 and observe the increase in model accuracy with a slight exception for UCM for $n=700$.  Figure~\ref{fig:pca_model_size_vs_accuracy} shows the effect of such scenarios (in experiments, we used fixed random states 3, 33, and 333 for PCA, RepeatedStratified K-Fold, and SVM respectively). This hyperparameter, that is, the number of PCA components, can be used as a knob for the accuracy vs PCA model size trade-off.
\begin{table}[]
\caption{Layer selection vs Accuracy trade-off}
\label{tab:ablation-study}
\centering
\adjustbox{width=0.75\linewidth}{

\begin{tabular}{cccccc}
\toprule[2pt]
Dataset & Layer 3 & Layer 6 & Layer 13 & Layer 16 & Accuracy\\
\midrule[1.5pt]
NWPU    & \cmark       & \xmark       & \xmark        & \xmark        & 71.56 $\pm$ 0.31     \\

NWPU    & \cmark       & \cmark       & \xmark        & \xmark        & 78.44 $\pm$ 0.22     \\
NWPU    & \cmark       & \cmark       & \cmark        & \xmark        & 87.5  $\pm$ 0.24     \\
NWPU    & \cmark       & \cmark       & \cmark        & \cmark        & 88.05  $\pm$ 0.18     \\
AID     & \cmark       & \xmark       & \xmark        & \xmark        & 82.07 $\pm$ 0.88    \\

AID     & \cmark       & \cmark       & \xmark        & \xmark        & 87.85 $\pm$ 0.56     \\
AID     & \cmark       & \cmark       & \cmark        & \xmark        & 93.64 $\pm$ 0.42     \\
AID     & \cmark       & \cmark       & \cmark        & \cmark        & 93.64 $\pm$ 0.48     \\
\bottomrule[2 pt]
\end{tabular}
\label{tab:modelselectionvsaccuracy}
}
\end{table}
\vspace{-1mm}
\subsection{Ablation Study}\label{sec:ablation}
\vspace{-3mm}
To show the effect of layer selection on accuracy, we add a layer from our selection algorithm from the top layer to the deeper layer. We observe an increase in accuracy in most of the cases. We experimented this with AID ($50\%$ training) and NWPU ($20\%$ training) datasets as shown in Table~\ref{tab:modelselectionvsaccuracy} using only LDA for dimension reduction and SVM for classification as the feature dimension changes every time. The result demonstrates the significance of our layer selection.

%We observe from Figure~\ref{fig:model-size} that the results with BN layers taken from blocks 3, 6, and 13 outperform the results when the BN layers are taken from blocks 9-12. To investigate further, we test all possible of 3 layers from this set with our selected 3 layers' features and report the results in Table~\ref{tab:all3layers} for the AID dataset with 50\% training. During this experiment, we have dropped the PCA layer and produced the results with LDA and SVM classification. 
%It is worth noted that our technique of selecting layers performs better compared to other layers in terms of accuracy. %In addition, from Figure~\ref{fig:alpha-aid}, the layer 10 is also significant in terms of significance, i.e. $\alpha$.   
\vspace{-2mm}
\section{Conclusion}
\vspace{-3mm}
In this paper, we have proposed a novel approach for the classification of aerial images that extracts and fuses features from selected layers of the pretrained MobileNetV2 model based on feature fusion strategy. We have shown that our constructed model achieves similar or superior classification accuracy although the size of the model is smaller compared to the earlier works and the training cost is very minimal. While we apply the ReLU-based feature fusion on MobileNetV2 for %a specific task at hand, that is, 
only the aerial scene classification, we recognize that the technique can be applied on other transfer learning tasks too.
%such as medical image classification. 
In the future, we intend to investigate further on this.

\bibliographystyle{IEEEbib}

\small{
\bibliography{ref}
}
\end{document}